\documentclass[preprint,aps]{revtex4}
\usepackage{amsmath}
\usepackage{graphicx}

\begin{document}

\title{Bounce Models in Brane Cosmology and a Gravitational Stability
Condition}
\author{Hongya Liu}
\affiliation{Department of Physics, Dalian University of Technology, Dalian 116024, P.R.
China}

\begin{abstract}
Five-dimensional cosmological models with two 3-branes and with a buck
cosmological constant are studied. It is found that for all the three cases (%
$\Lambda =0$, $\Lambda >0$, and $\Lambda <0$), the conventional space-time
singularity ``big bang'' could be replaced by a matter singularity ``big
bounce'', at which the ``size'' of the universe and the energy density are
finite while the pressure diverges, and across which the universe evolves
from a pre-existing contracting phase to the present expanding phase. It is
also found that for the $\Lambda >0$ case the brane solutions could give an
oscillating universe model in which the universe oscillates with each cosmic
cycle begins from a ``big bounce'' and ends to a ``big crunch'', with a
distinctive characteristic that in each subsequent cycle the universe
expands to a larger size and then contracts to a smaller (but non-zero)
size. By studying the gravitational force acted on a test particle in the
bulk, a gravitational stability condition is derived and then is used to
analyze those brane models. It predicts that if dark energy takes over
ordinary matter, particles on the brane may become unstable in the sense
that they may escape from our 4D-world and dissolve in the bulk due to the
repulsive force of dark energy.

-----------------


\textbf{E-mail}: hyliu@dlut.edu.cn
\end{abstract}

\maketitle

\section{INTRODUCTION}

In the brane-world scenarios, our conventional universe is a 3-brane
embedded in a higher dimensional space. While gravity can freely propagate
in all dimensions, the standard matter particles and forces are confined to
the 3-brane only [1]. In recent years, five-dimensional (5D) brane
cosmological models have received extensive studies [2-9]. It is noticed
that one of the many interesting features of brane models concerns the big
bang singularity: Due to the existence of extra dimensions, the conventional
big bang singularity could be removed. So the big bang is perhaps not the
beginning of time but a transition from a pre-existing phase of the universe
to the present expanding phase, and our universe may have existed for an
infinite time prior to the putative big bang. In the ekpyrotic model [10],
it was suggested that the universe was produced from a collision between our
brane and a bulk brane. In the cyclic model [11], the universe undergoes an
endless sequence of cosmic epochs that begin with a big bang and end in a
big crunch. In the bounce models [7-9,12-14], it was shown that the scale
factor could evolve across a finite (but non-zero) minimum which represents
a bounce (as opposed to a bang). The purpose of this paper is to study the
bounce property of brane models more generally.

In a previous paper [15], a five-dimensional big bounce cosmological
solution with a \textit{non-compact} fifth dimension was presented. By using
this solution as valid in the bulk, a global brane model is derived in Ref.
[9], in which the model has two 3-branes with the extra dimension
compactified on an $S_{1}/Z_{2}$ orbifold. This brane model is of the type
of Binetruy, Deffayet and Langlois [2,3] in which no cosmological constant
was introduced in the bulk. In this paper, we are going to generalize it by
adding a cosmological constant in the bulk, and then to study evolutions of
the models.

The plan of this paper is as follows. In Section II, we look for general
solutions for cosmological models with two 3-branes and for three cases with
$\Lambda =0$, $\Lambda >0$ and $\Lambda <0$, respectively. In Section III,
we study the gravitational force acted on a test particle in the vicinity of
a brane and derive a stability condition. In Section IV, we give several
simple exact solutions as an illustration, and study the global evolutions
and the stability of the brane models as well as the big bounce singularity.
Section V is a short discussion.

\section{GENERAL BRANE SOLUTIONS WITH A COSMOLOGICAL CONSTANT}

Let the 5D metric being
\begin{equation}
dS^{2}=B^{2}dt^{2}-A^{2}\left( \frac{dr^{2}}{1-kr^{2}}+r^{2}d\Omega
^{2}\right) -dy^{2}\;,  \label{5metr}
\end{equation}%
where $B=B(t_{,}y)$ and $A=A(t_{,}y)$ are two scale factors, $k$($=\pm 1$ or
$0$) is the 3D curvature index, and $d\Omega ^{2}\equiv (d\theta ^{2}+\sin
^{2}\theta d\phi ^{2})$. This metric describes 5D cosmological models with
spherical symmetry in 3D and with a static fifth dimension. Now we consider
brane models with the extra fifth dimension $y$ compactified on a small
circle for which the first brane is at $y=y_{1}=0$ and the second is at $%
y=y_{2}$ with $y_{2}>0$. The 5D Einstein equations read%
\begin{eqnarray}
G_{AB} &=&\kappa ^{2}\left( \Lambda g_{AB}+T_{AB}\right) \qquad ,  \notag \\
T_{AB} &=&\sum\limits_{i=1,2}\left[ \left( \rho _{i}+p_{i}\right) u^{\alpha
}u^{\beta }-p_{i}g^{\alpha \beta }\right] g_{\alpha A}g_{\beta B}\delta
\left( y-y_{i}\right) \;,  \label{5EinEqs}
\end{eqnarray}%
where upper case Latin letters denote 5D indices (0,1,2,3;5), lower case
Greek letters denote 4D indices (0,1,2,3), $\kappa ^{2}$ is the 5D
gravitational constant, $u^{\alpha }\equiv dx^{\alpha }/ds$ is the 4D
velocity in comoving coordinates with $ds$ being the 4D line-element and $%
u^{\alpha }\equiv (u^{0},0,0,0)$, and $\rho _{i}$ and $p_{i}$ are energy
density and pressure on the $i$-th brane, respectively. With use of the
metric (\ref{5metr}), the non-vanishing equations in (\ref{5EinEqs}) are
found to be%
\begin{eqnarray}
G_{00} &=&3\left( \frac{\dot{A}^{2}}{A^{2}}+k\frac{B^{2}}{A^{2}}\right)
-3B^{2}\left( \frac{A^{\prime \prime }}{A}+\frac{A^{\prime 2}}{A^{2}}\right)
\notag \\
&=&\kappa ^{2}B^{2}\left[ \Lambda +\sum\limits_{i=1,2}\rho _{i}\delta \left(
y-y_{i}\right) \right] \;,  \label{5EinG00}
\end{eqnarray}%
\begin{equation}
G_{05}=-3\left( \frac{\dot{A}^{\prime }}{A}-\frac{\dot{A}}{A}\frac{B^{\prime
}}{B}\right) =0\;,  \label{5EinG05}
\end{equation}%
\begin{equation}
G_{55}=-\frac{3}{B^{2}}\left[ \frac{\ddot{A}}{A}+\frac{\dot{A}}{A}\left(
\frac{\dot{A}}{A}-\frac{\dot{B}}{B}\right) +k\frac{B^{2}}{A^{2}}\right] +%
\frac{3A^{\prime }}{A}\left( \frac{A^{\prime }}{A}+\frac{B^{\prime }}{B}%
\right) =-\kappa ^{2}\Lambda \;,  \label{5EinG55}
\end{equation}%
\begin{eqnarray}
\left( 1-kr^{2}\right) G_{11} &=&r^{-2}G_{22}=r^{-2}\sin ^{-2}\theta G_{33}
\notag \\
&=&-\frac{A^{2}}{B^{2}}\left[ \frac{2}{A}+\frac{\dot{A}}{A}\left( \frac{\dot{%
A}}{A}-\frac{2\dot{B}}{B}\right) +k\frac{B^{2}}{A^{2}}\right]  \notag \\
&&+A^{2}\left[ \frac{B^{\prime \prime }}{B}+\frac{2A^{\prime \prime }}{A}+%
\frac{A^{\prime }}{A}\left( \frac{A^{\prime }}{A}+\frac{2B^{\prime }}{B}%
\right) \right]  \notag \\
&=&-\kappa ^{2}A^{2}\left[ \Lambda -\sum\limits_{i=1,2}p_{i}\delta \left(
y-y_{i}\right) \right] \;,  \label{5EinG11}
\end{eqnarray}%
where an overdot and a prime denote partial derivatives with respect to $t$
and $y$, respectively.

Note that there are different ways to express the solutions of above
equations (see, for example, [3]). Here we follow our previous work
[9,15,16] for convenience. We see that equation (\ref{5EinG05}) can be
integrated once, giving%
\begin{equation}
B=\frac{\dot{A}}{\mu (t)}\;,  \label{B=}
\end{equation}%
where $\mu (t)$ is an arbitrary function. Using this relation to eliminate $%
B $ from equation (\ref{5EinG00}), we obtain%
\begin{equation}
\left( A^{2}\right) ^{\prime \prime }+\frac{2}{3}\kappa ^{2}\left[ \Lambda
+\sum\limits_{i=1,2}\rho _{i}\delta \left( y-y_{i}\right) \right]
A^{2}=2\left( \mu ^{2}+k\right) \;.  \label{5EinG00b}
\end{equation}%
Also, with use of the relation (\ref{B=}), equation (\ref{5EinG55}) can then
be integrated, giving%
\begin{equation}
\left( \mu ^{2}+k-A^{\prime 2}\right) A^{2}=\frac{1}{6}\kappa ^{2}\Lambda
A^{4}+F\;,  \label{5EinG55b}
\end{equation}%
where $F$ is an integration ``constant'' which actually could be a function
of $y$.

In the bulk, the two equations (\ref{5EinG00b}) and (\ref{5EinG55b}) require
that $F=K=const$. So the Einstein equation in the \textit{bulk} is
\begin{equation}
\left( A^{2}\right) ^{\prime 2}+\frac{2}{3}\kappa ^{2}\Lambda A^{4}=4A^{2}%
\left[ \mu ^{2}(t)+k\right] -4K\;.  \label{5EinG55c}
\end{equation}%
Meanwhile, we can verify that equation (\ref{5EinG11}) is then satisfied
identically.

From a known exact bulk solution of the equation (\ref{5EinG55c}) we can
easily extend it from bulk to branes and obtain a global two-brane solution.
The $S_{1}/Z_{2}$\ symmetry requires that we firstly should write a bulk
solution from $g_{AB}(t,y)$ to $g_{AB}(t,\left| y\right| )$.\ Then,
according to Israel's jump conditions, the two scale factors $A(t,\left|
y\right| )$ and $B(t,\left| y\right| )$ are required to be continuous across
the branes localized in $y=y_{i}$. Their first derivatives with respect to $%
y $ can be discontinuous across the branes. And their second derivatives
with respect to $y$ can give a Dirac delta function. Thus the resulting
terms with a delta function appearing in the LHS of equations (\ref{5EinG00}%
) and (\ref{5EinG11}) must be matched with the corresponding terms
containing a delta function in the RHS of them in order to satisfy the field
equations. For $A_{i}$ we have, for instance,%
\begin{equation}
A_{i}^{\prime \prime }=\widehat{A_{i}^{\prime \prime }}+\left[ A^{\prime }%
\right] _{i}\delta (y-y_{i})\;,  \label{Ai''}
\end{equation}%
where $\widehat{A_{i}^{\prime \prime }}$ represents terms in $A_{i}^{\prime
\prime }$ that are not contributed to the delta function $\delta (y-y_{i})$,
and $\left[ A^{\prime }\right] _{i}$ is the jump in the first derivative
across $y=y_{i}$, defined by $\left[ A^{\prime }\right] _{i}\equiv A^{\prime
}(y_{i}^{+})-A^{\prime }(y_{i}^{-})$. The $Z_{2}$ reflection symmetry leads
to%
\begin{equation}
A^{\prime }(y_{i}^{-})=-A^{\prime }(y_{i}^{+})\;,\qquad \left[ A^{\prime }%
\right] _{i}\equiv 2A^{\prime }(y_{i}^{+})\;.  \label{[A']i}
\end{equation}%
Thus the Einstein equations (\ref{5EinG00}) and (\ref{5EinG11}) on the
branes give%
\begin{eqnarray}
\kappa ^{2}\,\rho _{i} &=&-\frac{3}{A_{i}}\left[ A^{\prime }\right] _{i}=-%
\frac{6A^{\prime }(y_{i}^{+})}{A_{i}}\;,  \notag \\
\kappa ^{2}\,p_{i} &=&\frac{1}{B_{i}}\left[ B^{\prime }\right] _{i}+\frac{2}{%
A_{i}}\left[ A^{\prime }\right] _{i}=\frac{2B^{\prime }(y_{i}^{+})}{B_{i}}+%
\frac{4A^{\prime }(y_{i}^{+})}{A_{i}}\;.  \label{rhoi,pi}
\end{eqnarray}%
Meanwhile, the 5D conservation law $T_{A\;;B}^{\;B}=0$ gives the usual 4D
equation of conservation on the branes:%
\begin{equation}
\dot{\rho}_{i}+3\left( \rho _{i}+p_{i}\right) \frac{\dot{A}_{i}}{A_{i}}=0\;.
\label{rhoi-cons}
\end{equation}

We also need to define the Hubble and deceleration parameters on the branes
appropriately. Be aware that the proper time on a given $y=$constant
hypersurface is defined by $d\tau =Bdt$, so, with use of the relation (\ref%
{B=}), the Hubble and deceleration parameters are defined as (see [9])%
\begin{equation}
H(t,y)\equiv \frac{1}{B}\frac{\dot{A}}{A}=\frac{\mu }{A}\;,\qquad q(t,y)=-%
\left[ \frac{A}{B}\frac{\partial }{\partial t}\left( \frac{\dot{A}}{B}%
\right) \right] \left( \frac{\dot{A}}{B}\right) ^{-2}=-\frac{A\dot{\mu}}{\mu
\dot{A}}\;.  \label{H,q}
\end{equation}%
So, on the branes, we have%
\begin{equation}
H_{i}=\frac{\mu }{A_{i}}\;,\qquad q_{i}=-\frac{A_{i}\dot{\mu}}{\mu \dot{A}%
_{i}}\;.  \label{Hi,qi}
\end{equation}%
Using (\ref{Hi,qi}) and (\ref{rhoi,pi}) in (\ref{5EinG55c}), we obtain%
\begin{equation}
H_{i}^{2}+\frac{k}{A_{i}^{2}}=\frac{\kappa ^{4}}{36}\rho _{i}^{2}+\frac{%
\kappa ^{2}}{6}\Lambda +\frac{K}{A_{i}^{4}}\;.  \label{Hi-rhoi}
\end{equation}%
Thus we recover the induced Friedmann equation on the branes that was
discussed and studied widely in literature [2-6,9,17]. In what follows we
consider the three cases for the brane models with $\Lambda =0$, $\Lambda >0$%
, and $\Lambda <0$, respectively.

\subsection{\qquad TYPE I BRANE MODELS: $\Lambda =0$}

For $\Lambda =0$, the general solutions of (\ref{5EinG55c}) are found to be
[9]
\begin{equation}
A^{2}=\left( \mu ^{2}+k\right) y^{2}-2\nu \left| y\right| +\frac{\nu ^{2}+K}{%
\mu ^{2}+k}\;.  \label{I-Sol}
\end{equation}%
This type of solutions contains two arbitrary functions $\mu (t)$ and $\nu
(t)$ and two constants $k$ and $K$. Differentiating this $A^{2}$ with
respect to $y$, we obtain
\begin{equation}
AA^{\prime }=\left( \mu ^{2}+k\right) y-\nu \frac{\partial \left| y\right| }{%
\partial y}\;.  \label{I-AA'}
\end{equation}%
Therefore we get
\begin{eqnarray}
A^{\prime }(0^{+}) &=&-A^{\prime }(0^{-})=-\frac{\nu }{A_{1}}\;,  \notag \\
A^{\prime }(y_{2}^{+}) &=&-A^{\prime }(y_{2}^{-})=-\frac{(\mu
^{2}+k)y_{2}-\nu }{A_{2}}\;,  \label{I-A'y+-}
\end{eqnarray}%
where the first brane is at $y=y_{1}=0$ and the second is at $y=y_{2}>0$.
Then, from (\ref{B=}) we get
\begin{eqnarray}
B^{\prime }(0^{+}) &=&-B^{\prime }(0^{-})=-\frac{1}{\mu }\frac{\partial }{%
\partial t}\left( \frac{\nu }{A_{1}}\right) \;,  \notag \\
B^{\prime }(y_{2}^{+}) &=&-B^{\prime }(y_{2}^{-})=-\frac{1}{\mu }\frac{%
\partial }{\partial t}\left[ \frac{(\mu ^{2}+k)y_{2}-\nu }{A_{2}}\right] \;.
\label{I-B'y+-}
\end{eqnarray}%
So equation (\ref{rhoi,pi}) gives
\begin{eqnarray}
\kappa ^{2}\,\rho _{1} &=&\frac{6\nu }{A_{1}^{2}}\;,  \notag \\
\kappa ^{2}\,p_{1} &=&-\frac{2}{\dot{A}_{1}}\frac{\partial }{\partial t}%
\left( \frac{\nu }{A_{1}}\right) -\frac{4\nu }{A_{1}^{2}}\;,
\label{I-rho1p1}
\end{eqnarray}%
and
\begin{eqnarray}
\kappa ^{2}\,\rho _{2} &=&\frac{6}{A_{2}}\left( \frac{(\mu ^{2}+k)y_{2}-\nu
}{A_{2}}\right) \;,  \notag \\
\kappa ^{2}\,p_{2} &=&-\frac{2}{\dot{A}_{2}}\frac{\partial }{\partial t}%
\left( \frac{(\mu ^{2}+k)y_{2}-\nu }{A_{2}}\right) -\frac{4}{A_{2}}\left(
\frac{(\mu ^{2}+k)y_{2}-\nu }{A_{2}}\right) \;.  \label{I-rho2p2}
\end{eqnarray}

\subsection{TYPE II BRANE MODELS: $\Lambda >0$}

For $\Lambda >0$, the general solutions of (\ref{5EinG55c}) can easily be
found. Then, by changing $y$ to $\left| y\right| $, we obtain%
\begin{eqnarray}
A^{2} &=&a(t)\cos \left( \theta (t)+L^{-1}\left| y\right| \right)
+2L^{2}\left( \mu ^{2}+k\right) \;,  \notag \\
a(t) &=&2L\sqrt{L^{2}\left( \mu ^{2}+k\right) ^{2}-K}\;,\qquad L^{-1}\equiv
\sqrt{\frac{2\kappa ^{2}\Lambda }{3}}\;,  \label{II-Sol}
\end{eqnarray}%
where $\mu (t)$ and $\theta (t)$ are two arbitrary functions. By
differentiating $A^{2}$ with respect to $y$, we obtain%
\begin{equation}
2AA^{\prime }=-L^{-1}a(t)\sin \left( \theta +L^{-1}\left| y\right| \right)
\frac{\partial \left| y\right| }{\partial y}\;.  \label{II-AA'}
\end{equation}%
So we have%
\begin{eqnarray}
A^{\prime }(0^{+}) &=&-A^{\prime }(0^{-})=-\frac{a\sin \theta }{2LA_{1}}%
\;,\quad  \notag \\
A^{\prime }(y_{2}^{+}) &=&-A^{\prime }(y_{2}^{-})=\frac{a\sin \left( \theta
+L^{-1}y_{2}\right) }{2LA_{2}}\;,  \label{II-A'y+-}
\end{eqnarray}%
Using these and the relation $B=\dot{A}/\mu $, we get%
\begin{eqnarray}
B^{\prime }(0^{+}) &=&-B^{\prime }(0^{-})=-\frac{1}{2L\mu }\frac{\partial }{%
\partial t}\left( \frac{a\sin \theta }{A_{1}}\right) \;,\qquad  \notag \\
B^{\prime }(y_{2}^{+}) &=&-B^{\prime }(y_{2}^{-})=\frac{1}{2L\mu }\frac{%
\partial }{\partial t}\left( \frac{a\sin \left( \theta +L^{-1}y_{2}\right) }{%
A_{2}}\right) \;,\;  \label{II-B'y+-}
\end{eqnarray}%
So equation (\ref{rhoi,pi}) gives%
\begin{eqnarray}
\kappa ^{2}\,\rho _{1} &=&\frac{3a\sin \theta }{LA_{1}^{2}}\;,  \notag \\
\kappa ^{2}\,p_{1} &=&-\frac{1}{L\dot{A}_{1}}\frac{\partial }{\partial t}%
\left( \frac{a\sin \theta }{A_{1}}\right) -\frac{2a\sin \theta }{LA_{1}^{2}}%
\;,  \label{II-rho1p1}
\end{eqnarray}%
and%
\begin{eqnarray}
\kappa ^{2}\,\rho _{2} &=&-\frac{3a\sin \left( \theta +L^{-1}y_{2}\right) }{%
LA_{2}^{2}}\;,  \notag \\
\kappa ^{2}\,p_{2} &=&\frac{1}{L\dot{A}_{2}}\frac{\partial }{\partial t}%
\left( \frac{a\sin \left( \theta +L^{-1}y_{2}\right) }{A_{2}}\right) +\frac{%
2a\sin \left( \theta +L^{-1}y_{2}\right) }{LA_{2}^{2}}\;.  \label{II-rho2p2}
\end{eqnarray}

\subsection{TYPE III BRANE MODELS: $\Lambda <0$}

For $\Lambda <0$, the general solutions of (\ref{5EinG55c}) are found to be%
\begin{eqnarray}
A^{2} &=&a(t)\cosh \left( \xi (t)+L^{-1}\left| y\right| \right)
-2L^{2}\left( \mu ^{2}+k\right) \;,  \notag \\
a(t) &=&2L\sqrt{L^{2}\left( \mu ^{2}+k\right) ^{2}+K}\;,\qquad L^{-1}\equiv
\sqrt{-\frac{2}{3}\kappa ^{2}\Lambda }\;,  \label{III-Sol}
\end{eqnarray}%
where $k$ and $K$ and $L$ are constants, $\mu =\mu (t)$ and $\xi =\xi (t)$
are two arbitrary functions. Differentiating $A^{2}$ with respect to $y$, we
obtain

\begin{equation}
2AA^{\prime }=L^{-1}a(t)\sinh \left( \xi +L^{-1}\left| y\right| \right)
\frac{\partial \left| y\right| }{\partial y}\;.  \label{III-AA'}
\end{equation}%
So we find%
\begin{eqnarray}
A^{\prime }(0^{+}) &=&-A^{\prime }(0^{-})=\frac{a\sinh \xi }{2LA_{1}}\;,
\notag \\
A^{\prime }(y_{2}^{+}) &=&-A^{\prime }(y_{2}^{-})=-\frac{a\sinh \left( \xi
+L^{-1}y_{2}\right) }{2LA_{2}}\;,  \label{III-A'y+-}
\end{eqnarray}%
Then, using these and the relation $B=\dot{A}/\mu $, we find%
\begin{eqnarray}
B^{\prime }(0^{+}) &=&-B^{\prime }(0^{-})=\frac{1}{2L\mu }\frac{\partial }{%
\partial t}\left( \frac{a\sinh \xi }{A_{1}}\right) \;,\qquad  \notag \\
B^{\prime }(y_{2}^{+}) &=&-B^{\prime }(y_{2}^{-})=-\frac{1}{2L\mu }\frac{%
\partial }{\partial t}\left( \frac{a\sinh \left( \xi +L^{-1}y_{2}\right) }{%
A_{2}}\right) \;.\;  \label{III-B'y+-}
\end{eqnarray}%
So the equation (\ref{rhoi,pi}) gives%
\begin{eqnarray}
\kappa ^{2}\,\rho _{1} &=&-\frac{3a\sinh \xi }{LA_{1}^{2}}\;,  \notag \\
\kappa ^{2}\,p_{1} &=&\frac{1}{L\dot{A}_{1}}\frac{\partial }{\partial t}%
\left( \frac{a\sinh \xi }{A_{1}}\right) +\frac{2a\sinh \xi }{LA_{1}^{2}}\;,
\label{III-rho1p1}
\end{eqnarray}%
and%
\begin{eqnarray}
\kappa ^{2}\,\rho _{2} &=&\frac{3a\sinh \left( \xi +L^{-1}y_{2}\right) }{%
LA_{2}^{2}}\;,  \notag \\
\kappa ^{2}\,p_{2} &=&-\frac{1}{L\dot{A}_{2}}\frac{\partial }{\partial t}%
\left( \frac{a\sinh \left( \xi +L^{-1}y_{2}\right) }{A_{2}}\right) -\frac{%
2a\sinh \left( \xi +L^{-1}y_{2}\right) }{LA_{2}^{2}}\;.  \label{III-rho2p2}
\end{eqnarray}

\section{A GRAVITATIONAL STABILITY CONDITION FOR PARTICLES ON BRANES}

Generally speaking, brane models require the fifth dimension to be
compactified to a small size $y_{2}$. So a stable brane model should have a
stable size.\ The well-known Goldberger-Wise mechanism [22] was to use a
scalar field in the bulk to stabilize the size of the fifth dimension. In
this way, the model has the tendency to recover it's size after a
perturbation. Here, a perturbation means a small change for the size.

Here we wish to say that even if the size of the fifth dimension is somehow
stabilized, one still can ask question whether particles on the brane may
leave the brane and escape into the bulk. Arkani-Hamed et al [1] pointed out
that in sufficiently hard collisions, particles on the brane can acquire
momentum in the extra dimensions and escape from our 4D world, carrying away
energy. If this happens continuously, it will cause another kind of
instability problem. Note that here the instability just means brane
particles are not in a stable position along the fifth dimension. Now a
natural question is that once a particle left the brane, will it escape in
the bulk forever or return to our brane again?

To answer this question we noticed that the bulk discussed in this paper
only contains a cosmological constant term $\Lambda $\ and so is empty.
Therefore, it is reasonable to expect that particles inside the bulk should
obey the 5D geodesic equation%
\begin{equation}
\frac{d^{2}x^{A}}{dS^{2}}+\Gamma _{BC}^{A}\frac{dx^{B}}{dS}\frac{dx^{C}}{dS}%
=0\;,  \label{5DGeo}
\end{equation}%
which was used previously [18] for a similar purpose as here. From this
equation, a 5D gravitational force acted on the bulk test particle can be
defined as%
\begin{equation}
F^{A}=-\Gamma _{BC}^{A}\frac{dx^{B}}{dS}\frac{dx^{C}}{dS}\;.  \label{5D-F}
\end{equation}%
For simplicity, we assume that the particle is temporary at rest in the
bulk. Then, along the fifth dimension, (\ref{5D-F}) gives%
\begin{equation}
F^{5}=-\frac{B^{\prime }}{B}\;.  \label{F^5}
\end{equation}%
In the vicinity of the $i$-th brane, we use the general results (\ref%
{rhoi,pi}) and (\ref{[A']i}) in (\ref{F^5}). Then we obtain%
\begin{equation}
F^{5}(y_{i}^{+})=-F^{5}(y_{i}^{-})=-\frac{\kappa ^{2}}{3}\,\left( \rho _{i}+%
\frac{3}{2}p_{i}\right) \;.  \label{F^5(y+-)}
\end{equation}%
If this force $F^{5}$ is attractive, i.e., $F^{5}(y_{i}^{+})<0$ and $%
F^{5}(y_{i}^{-})>0$, the particle would be ``pulled'' back to our brane.
Thus we obtain a new kind of stability condition for the $i$-th brane as%
\begin{equation}
\rho _{i}+\frac{3}{2}p_{i}>0\;.  \label{Stab-Cond}
\end{equation}

This is a reasonable condition which holds for ordinary matters including
dark matter. However, if the brane matter contains a dark energy component
such as a cosmological constant or quintessence, the condition (\ref%
{Stab-Cond}) may or may not hold, depending on how much the dark energy is
contained on the brane. For example, we let%
\begin{equation}
\rho _{i}=\bar{\rho}_{i}+\bar{\lambda}_{i}\;,\qquad p_{i}=\bar{p}_{i}-\bar{%
\lambda}_{i}\;,  \label{rhoi-bar}
\end{equation}%
where $\bar{\lambda}_{i}$ is a cosmological term on the $i$-th brane. Then
the condition (\ref{Stab-Cond}) becomes%
\begin{equation}
\bar{\rho}_{i}+\frac{3}{2}\bar{p}_{i}>\frac{1}{2}\bar{\lambda}_{i}\;.
\label{Stab-Cond-bar}
\end{equation}%
Suppose initially condition (\ref{Stab-Cond-bar}) holds, then, as the
universe expands, both $\bar{\rho}_{i}$ and $\bar{p}_{i}$ decrease while $%
\bar{\lambda}_{i}$ keeps unchanged. So gradually the universe will enter in
an unstable stage in which particles and energy on the brane may escape in
the bulk and the 4D conservation law of energy (\ref{rhoi-cons}) may not
hold anymore. Recent observations [19,20] reveal that presently we are
living in an accelerating stage dominated by a dark energy term such as $%
\bar{\lambda}_{i}$\ in (\ref{Stab-Cond-bar}). Thus we see that both the
acceleration of our universe and the unstable nature of the brane particles
could be explained as due to the same repulsive force of dark energy. In the
following section we will show that some brane models do not satisfy the
condition (\ref{Stab-Cond}).

\section{SOME SIMPLE BIG BOUNCE BRANE MODELS}

In Section II we have obtained three types of global brane solutions
corresponding to $\Lambda =0$, $\Lambda >0$ and $\Lambda <0$, respectively.
Each type contains two arbitrary functions: $\mu (t)$ and $\nu (t)$ for type
I, $\mu (t)$ and $\theta (t)$ for type II, and $\mu (t)$ and $\xi (t)$ for
type III. All the two scale factors $A(t,y)$ and $B(t,y)$ and the densities $%
\rho _{i}$ and $p_{i}$ on the branes are expressed via the two functions.
From the relation $B=\dot{A}/\mu $ and metric (\ref{5metr}) we see that the
form of $Bdt$ is invariant under an arbitrary coordinate transformation $%
t\rightarrow \widetilde{t}(t)$. This freedom could be used to fix one of the
two arbitrary functions. Another freedom may correspond, as is in the
standard general relativity, to the unspecified equation of state of matter.
So, generally speaking, if the matter content on the first brane is known,
then the two arbitrary functions could be fixed. Then the whole solutions
could be fixed too. Then we will know the matter content on the second
brane. Therefore, the brane solutions obtained in Sec. II are quite general.

To compare these cosmological solutions with observations, we need know
clearly the matter content on our brane. This may need careful analysis and
might be complicated mathematically, and we are not going to do it here in
this paper. However, as an illustration, we will pick up several simple
exact models in the following and to exhibit typical features of brane
cosmology.

\subsection{A SIMPLE TWO-BRANE MODEL WITH $\Lambda =0$}

Consider the $\Lambda =0$\ brane solutions (\ref{I-Sol}), for which we choose%
\begin{equation}
k=0\;,\qquad K=1\;,\qquad \nu (t)=t_{b}/t\;,\qquad \mu (t)=\left( 2t\right)
^{-1/2}\quad ,  \label{I-spec}
\end{equation}%
where $t_{b}>0$ is a constant. In this way, the solution (\ref{I-Sol})
becomes
\begin{eqnarray}
A^{2} &=&2t\left[ 1+\left( \frac{\left| y\right| -2t_{b}}{2t}\right) ^{2}%
\right] \;,  \notag \\
B^{2} &=&\left[ 1-\left( \frac{\left| y\right| -2t_{b}}{2t}\right) ^{2}%
\right] ^{2}\left[ 1+\left( \frac{\left| y\right| -2t_{b}}{2t}\right) ^{2}%
\right] ^{-1}\;.  \label{I-ABspec}
\end{eqnarray}%
So on the first brane we have
\begin{eqnarray}
A_{1}^{2} &=&2t\left[ 1+\left( \frac{t_{b}}{t}\right) ^{2}\right] \;,\qquad
B_{1}^{2}=\left[ 1-\left( \frac{t_{b}}{t}\right) ^{2}\right] ^{2}\left[
1+\left( \frac{t_{b}}{t}\right) ^{2}\right] ^{-1}\;,  \notag \\
\kappa ^{2}\rho _{1} &=&\frac{3t_{b}}{t^{2}+t_{b}^{2}}\;,\qquad \kappa
^{2}p_{1}=\frac{2t_{b}}{t^{2}-t_{b}^{2}}-\frac{t_{b}}{t^{2}+t_{b}^{2}}\;.
\label{I-rho1spec}
\end{eqnarray}%
On the second brane we have
\begin{eqnarray}
A_{2}^{2} &=&2t\left[ 1+\left( \frac{y_{2}-2t_{b}}{2t}\right) ^{2}\right]
\;,\quad B_{2}^{2}=\left[ 1-\left( \frac{y_{2}-2t_{b}}{2t}\right) ^{2}\right]
^{2}\left[ 1+\left( \frac{y_{2}-2t_{b}}{2t}\right) ^{2}\right] ^{-1}\;
\notag \\
\kappa ^{2}\,\rho _{2} &=&\frac{6\left( y_{2}-2t_{b}\right) }{4t^{2}+\left(
y_{2}-2t_{b}\right) ^{2}}\;,\quad \kappa ^{2}\,p_{2}=\frac{4\left(
y_{2}-2t_{b}\right) }{4t^{2}-\left( y_{2}-2t_{b}\right) ^{2}}-\frac{2\left(
y_{2}-2t_{b}\right) }{4t^{2}+\left( y_{2}-2t_{b}\right) ^{2}}\;.
\label{I-rho2spec}
\end{eqnarray}

It is easy to see for the first brane that there is a critical time $%
t=t_{b}>0$ at which the scale factor $A_{1}(t)$ reaches to a non-zero
minimum $2\sqrt{t_{b}}$. After this critical time, $A_{1}(t)$\ increases,
implying the universe is expanding. For $t\gg t_{b}$ , we have $%
A_{1}(t)\rightarrow \sqrt{2t}$, $B_{1}(t)\rightarrow 1$, $\rho
_{1}\rightarrow 3p_{1}$, and the universe evolves as if in the
radiation-dominated standard Friedmann model. Before this critical time, the
universe contracts from infinity. It was shown [9,21] that as the coordinate
time $t$ varies from zero to $t_{b}$ , the proper time $\tau $ varies from $%
-\infty $ to $\tau _{b}$ , where $\tau _{b}$ is a finite constant and
corresponds to $t_{b}$ . Thus we can call $t_{b}$ as to represent a big
bounce (as opposed to a big bang), and before this bounce the universe has
existed for an infinitely long time. Meanwhile, rather than the big bang
singularity, which should correspond to $A_{1}(t)=0$, this big bounce
singularity corresponds to $B_{1}(t)=0$. From the solution (\ref{I-rho1spec}%
) we can also see that as $t$ varies across $t_{b}$ , the energy density $%
\rho _{1}$ remains finite while the pressure $p_{1}$ changes from
$-\infty $ to $+\infty $ . This implies a phase transition
happened at the bounce. Here we plot the evolution of the scale
factor $A_{1}(t)$ with $t_{b}=1$\ in Figure 1.

\begin{figure}[tbp]
\centering\includegraphics[width=2.5in,height=2.5in]{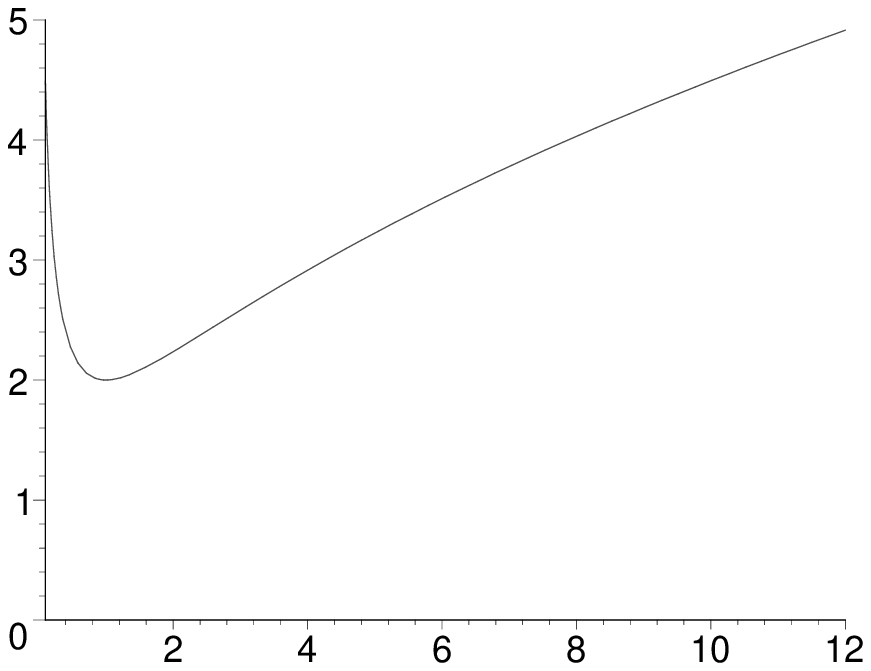}
\caption{Evolution of the scale factor
$A_{1}(t)=\protect\sqrt{2\left( t+1/t\right) }$ in a $\Lambda =0 $
brane cosmological model.}
\end{figure}

The evolution of the second brane is similar as the first one with the
bounce point at $t=t_{b2}\equiv \left| t_{b}-y_{2}/2\right| $. From equation
(\ref{I-rho2spec}) we see that whether the energy density on the second
brane is positive or negative depends on the ``size'' of the fifth dimension
$y_{2}$. If $y_{2}>2t_{b}$, we have $\rho _{2}>0$. If $y_{2}<2t_{b}$, we
have $\rho _{2}<0$.

Now let us look at the stability condition (\ref{Stab-Cond}). Using (\ref%
{I-rho1spec}) and (\ref{I-rho2spec}) in (\ref{Stab-Cond}), we obtain%
\begin{equation}
\kappa ^{2}\left( \rho _{1}+\frac{3}{2}p_{1}\right) =\frac{3t_{b}}{%
t^{2}-t_{b}^{2}}+\frac{3t_{b}}{2\left( t^{2}+t_{b}^{2}\right) }\;,
\label{I-stab-1}
\end{equation}%
\begin{equation}
\kappa ^{2}\left( \rho _{2}+\frac{3}{2}p_{2}\right) =\frac{6\left(
y_{2}-2t_{b}\right) }{4t^{2}-\left( y_{2}-2t_{b}\right) ^{2}}+\frac{3\left(
y_{2}-2t_{b}\right) }{4t^{2}+\left( y_{2}-2t_{b}\right) ^{2}}\;.
\label{I-stab-2}
\end{equation}%
For the first brane we find that $\rho _{1}+\frac{3}{2}p_{1}>0$ for $t>t_{b}$%
. So, in the expanding stage (after the bounce), the first brane is stable.
There are two cases for the second brane. If $y_{2}>2t_{b}$, we have $\rho
_{2}>0$ and $\rho _{2}+\frac{3}{2}p_{2}>0$ in the stage $t>t_{b2}$ where $%
t_{b2}\equiv \left| y_{2}-2t_{b}\right| $ represents the bounce time. So
after the bounce this brane is stable. If $y_{2}<2t_{b}$, we have $\rho
_{2}<0$ and $\rho _{2}+\frac{3}{2}p_{2}<0$ in the stage $t>t_{b2}$. So after
the bounce this brane is unstable. Thus, to obtain a model with two stable
branes, we must have $y_{2}>2t_{b}$. An interesting special case is $%
y_{2}=4t_{b}$ for which we have $\rho _{2}=\rho _{1}>0$ and the model is
completely symmetric and particles on the branes are stable.

\subsection{AN OSCILLATING UNIVERSE MODEL WITH $\Lambda >0$}

The notion of an oscillatory universe can be traced back to 1930's and has
continued to attract interest [23-25,11]. Tolman [23] discussed it within
the framework of general relativity assuming a closed universe ($k=+1$) in
which the universe undergoes a sequence of cycles of expansion and
contraction. Dicke and Peebles [24] restudied Tolman's model and pointed out
that an oscillating universe could provide an escape from some of the
cosmological problems such as the horizon and homogeneity problems. ``%
\textit{As the universe ages}'', they wrote, ``\textit{more and more of it
becomes visible that earlier was beyond the horizon and presumably causally
disconnected from us .... How then are we to understand the remarkable
familiarity of the objects just appearing on the horizon? Perhaps by tracing
the evolution back through the big bang to an earlier collapsing phase}''.
However, Tolman's oscillatory models were constrained by having to pass
through the big bang singularity in which the energy density and temperature
diverge. Therefore, Dicke and Peebles expected that ``\textit{some future
and better theory might show that the collapse of the universe would lead to
a `bounce' instead of a singularity}''.

Now let us consider the type II ($\Lambda >0$) brane solutions. As we
discussed at the beginning of this section that the metric form of $Bdt$ is
invariant under an arbitrary coordinate transformation $t\rightarrow
\widetilde{t}(t)$. Meanwhile, the scale factor $A(t,y)$ in (\ref{II-Sol})
contains two arbitrary functions $\mu (t)$ and $\theta (t)$. So, by choosing
the time coordinate properly, we can set $\theta (t)=-\pi t$ without loss of
generality. So the cosine term in the solution leads generally to an
oscillating universe model. For illustration, we choose

\begin{eqnarray}
\mu ^{2}(t) &=&t,\qquad \theta (t)=-\pi t\;,  \notag \\
L &=&K=1,\qquad k=0\;.  \label{II-spec}
\end{eqnarray}%
Then%
\begin{eqnarray}
A^{2} &=&2t+2\sqrt{t^{2}-1}\cos \left( \pi t-\left| y\right| \right)  \notag
\\
a(t) &=&2\sqrt{t^{2}-1}\;,\qquad \frac{2\kappa ^{2}\Lambda }{3}=1\;,
\label{II-Aspec}
\end{eqnarray}%
So on the branes we have%
\begin{equation}
A_{i}^{2}=2t+2\sqrt{t^{2}-1}\cos \left( \pi t-y_{i}\right) \;.
\label{II-Aispec}
\end{equation}%
We plot the evolution of the scale factor $A_{1}(t)$ in Figure 2. From this
figure we see clearly that the big bang spacetime singularity of the
standard cosmology is replaced here by a series of smooth big bounces at $%
t=t_{b}\approx 1$, $3$, $5$, $7$, ....

\begin{figure}[tbp]
\centering\includegraphics[width=2.5in,height=2.5in]{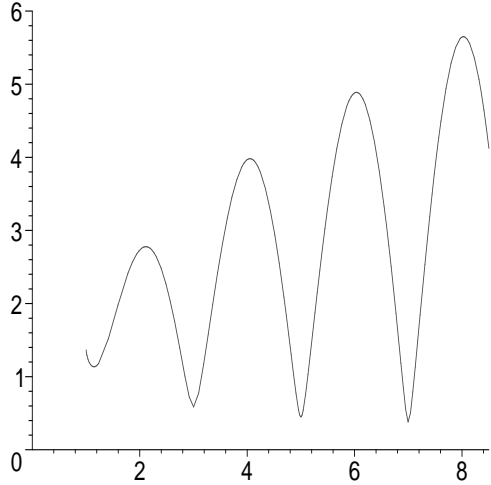}
\caption{Evolution of the scale factor $A_{1}(t)=\protect\sqrt{%
2t+2\protect\sqrt{t^{2}-1}\cos \left( \protect\pi t\right) }$ in a
$\Lambda
>0$ oscillatory brane cosmological model}
\end{figure}

Another typical feature of Figure 2 is that there is a ``beginning of time''
at $t=1$ in the model and the oscillating amplitudes of the followed cycles
increase monotonously. This reminds us of the well-known entropy problem of
the old oscillatory universe model. Tolman [23] pointed out that if the
total entropy of the universe can only increase, then, in an oscillatory
model, the entropy produced during one cycle would add to the entropy
produced in the next cycle, causing the oscillating amplitude of each cycle
to be larger than the one before it. Extrapolating backward in time, the
universe would have to be finite cycles old. More discussions about the
entropy problem can be found in literature [24,11]. We see that our
oscillating universe model described by equation (\ref{II-Aispec}) and
Figure 2 coincides with Tolman's description perfectly and, therefore, could
provide a realized framework to discuss the entropy problem.

The two equations (\ref{II-rho1p1}) and (\ref{II-rho2p2}) become%
\begin{eqnarray}
\kappa ^{2}\,\rho _{1} &=&-\frac{6\sqrt{t^{2}-1}\sin \pi t}{A_{1}^{2}}\quad ,
\notag \\
\kappa ^{2}\,p_{1} &=&\frac{2}{\dot{A}_{1}}\frac{\partial }{\partial t}%
\left( \frac{\sqrt{t^{2}-1}\sin \pi t}{A_{1}}\right) +\frac{4\sqrt{t^{2}-1}%
\sin \pi t}{A_{1}^{2}}\;,  \label{II-rho1spec}
\end{eqnarray}%
and%
\begin{eqnarray}
\kappa ^{2}\,\rho _{2} &=&\frac{6\sqrt{t^{2}-1}\sin \left( \pi
t-y_{2}\right) }{A_{2}^{2}}\;,  \notag \\
\kappa ^{2}\,p_{2} &=&-\frac{2}{\dot{A}_{2}}\frac{\partial }{\partial t}%
\left[ \frac{\sqrt{t^{2}-1}\sin \left( \pi t-y_{2}\right) }{A_{2}}\right] -%
\frac{4\sqrt{t^{2}-1}\sin \left( \pi t-y_{2}\right) }{A_{2}^{2}}\;.
\label{II-rho2spec}
\end{eqnarray}

So the energy density on the branes changes sign periodically with time,
implying that the universe may have a negative energy density. This is an
unusual feature of the oscillatory brane model. It is reasonable to assume
that the size the fifth dimension is much smaller than the period of each
cycle of the universe, i.e., $y_{2}\ll 2\pi $. Then from equations (\ref%
{II-Aispec}) - (\ref{II-rho2spec}) we find $A_{2}(t)\approx $ $A_{1}(t)$, $%
\rho _{2}\approx -\rho _{1}$ and $p_{2}\approx -p_{1}$. So if our brane has
a positive energy density at present stage of the universe, the other brane
would have a negative energy density. Applying this in the stability
condition (\ref{Stab-Cond}), we find that $\left( \rho _{2}+\frac{3}{2}%
p_{2}\right) \approx -\left( \rho _{1}+\frac{3}{2}p_{1}\right) $. This means
that if one of the two branes is at a stable stage, the other one must be at
an unstable stage. As a whole, we conclude that brane particles in this
oscillating model are not stable.

\subsection{A SIMPLE UNIVERSE MODEL WITH $\Lambda <0$}

For this type of brane solutions (\ref{III-Sol}), we take a similar choice
as in equations (\ref{II-spec}):%
\begin{eqnarray}
\mu ^{2}(t) &=&t,\qquad \xi (t)=-t\;,  \notag \\
L &=&K=1,\qquad k=0\;.  \label{III-spec}
\end{eqnarray}%
Then we obtain%
\begin{eqnarray}
A^{2} &=&2\sqrt{t^{2}+1}\cosh \left( t-\left| y\right| \right) -2t\;,  \notag
\\
A_{i}^{2} &=&2\sqrt{t^{2}+1}\cosh \left( t-y_{i}\right) -2t\;.
\label{III-Aispec}
\end{eqnarray}%
We plot the evolution of $A_{1}(t)$ of the first brane in Figure 3, from
which we see that $A_{1}(t)$\ reaches to its minimum at the bounce point $%
t=t_{b}\approx 0.5$. Before the bounce, the universe contracts
from infinity; after the bounce, the universe expands to infinity
again.

\begin{figure}[tbp]
\centering\includegraphics[width=2.5in,height=2.5in]{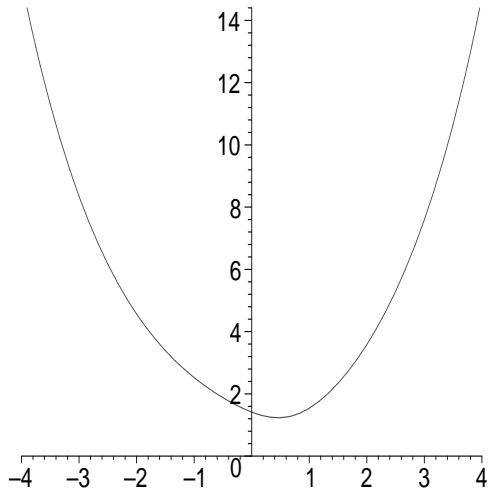}
\caption{Evolution of the scale
factor $A_{1}(t)=\protect\sqrt{2\protect\sqrt{t^{2}+1}\cosh t-2t}$ in a $%
\Lambda <0$ brane cosmological model.}
\end{figure}

From equations (\ref{III-rho1p1}) and (\ref{III-rho2p2}) we obtain%
\begin{eqnarray}
\kappa ^{2}\,\rho _{1} &=&\frac{6\sqrt{t^{2}+1}\sinh t}{A_{1}^{2}}\;,  \notag
\\
\kappa ^{2}\,p_{1} &=&-\frac{2}{\dot{A}_{1}}\frac{\partial }{\partial t}%
\left( \frac{\sqrt{t^{2}+1}\sinh t}{A_{1}}\right) -\frac{4\sqrt{t^{2}+1}%
\sinh t}{A_{1}^{2}}\;,  \label{III-rho1spec}
\end{eqnarray}%
and%
\begin{eqnarray}
\kappa ^{2}\,\rho _{2} &=&-\frac{6\sqrt{t^{2}+1}\sinh \left( t-y_{2}\right)
}{A_{2}^{2}}\;,  \notag \\
\kappa ^{2}\,p_{2} &=&\frac{2}{\dot{A}_{2}}\frac{\partial }{\partial t}%
\left( \frac{\sqrt{t^{2}+1}\sinh \left( t-y_{2}\right) }{A_{2}}\right) +%
\frac{4\sqrt{t^{2}+1}\sinh \left( t-y_{2}\right) }{A_{2}^{2}}\;.
\label{III-rho2spec}
\end{eqnarray}%
Thus the energy density is positive on the first brane and negative on the
second brane. If we assume $\left| t\right| \gg y_{2}$, then we have $\rho
_{2}\approx -\rho _{1}$ and $p_{2}\approx -p_{1}$. So probably we are living
on the first brane. As for the stability condition (\ref{Stab-Cond}), we
also have $\left( \rho _{2}+\frac{3}{2}p_{2}\right) \approx -\left( \rho
_{1}+\frac{3}{2}p_{1}\right) $ as in the above oscillating model. So
particles in this model are also unstable.

\section{DISCUSSION}

In this paper we have derived, in Sec. II, a class of five-dimensional
cosmological solutions with two 3-branes and with the fifth dimension being
static and compactified on a small circle. The bulk contains only a
cosmological constant $\Lambda $. It is found that for all the three cases
of $\Lambda $ ($\Lambda =0$, $\Lambda >0$, $\Lambda <0$) the solutions
contain two arbitrary functions of time. One of these two freedoms might be
explained as due to the unspecified time coordinate in the 5D metric (\ref%
{5metr}), leaving another to account for various contents of the cosmic
matter. In Section III we have used the 5D geodesic equations to study the
gravitational force acted on a test particle in the vicinity of a brane.
This force could be interpreted as generated by matters on the brane. By
requiring this force being attractive and so to grip particles from escaping
into the bulk, we have derived a physically reasonable gravitational
stability condition as given in equation (\ref{Stab-Cond}). For illustration
and for simplicity we presented three simple exact models in Section IV by
choosing the two arbitrary functions properly. From these simple models we
found that the conventional space-time singularity ``big bang'' could be
replaced in brane models by a matter singularity ``big bounce'' at which the
``size'' of the 3D space is finite and the energy density does not diverge,
while the pressure diverges. This enable us to expect that in brane
cosmological models the ``history'' of our universe could be traced back
across the big bounce to a pre-existing phase. This is clearly of great
interest and deserve more studies. The stability of brane particles of these
simple models are also studied.

Here we want to discuss more about the oscillating universe models given in
equations (\ref{II-Aspec}) - (\ref{II-rho2spec}). As pointed out by Tolman
that an oscillatory model could resolve the horizon and the homogeneity
problems. However, the main difficulty of Tolman's oscillatory model is
having to pass through the big bang space-time singularity. Now brane models
could remove the big bang singularity in a satisfactory way and thus rescued
Tolman's old model. Meanwhile, Tolman's entropy problem also get resolved.

We should emphasis that the present work is exploratory. Be aware that the
general brane solutions given in this paper contain two arbitrary functions.
This would enable us to discuss more observations such as the acceleration
of the universe [23]. We leave these studies in the future.

\begin{acknowledgments}
The author thanks Guowen Peng and Lixin Xu for useful discussions. This work
was supported by NSF of P. R. China under Grant 10273004.
\end{acknowledgments}

\end{document}